\documentclass[aip,pop,reprint]{revtex4-1}
\pdfoutput=1
\usepackage[utf8]{inputenc}
\usepackage[T1]{fontenc}
\usepackage{mathtools}
\usepackage{amsfonts}
\usepackage{amssymb}
\usepackage{graphicx}
\usepackage{siunitx}
\usepackage{lmodern}
\usepackage{physics}
\usepackage{microtype}
\usepackage{hyperref}
\hypersetup{
    linkcolor=blue,
    citecolor=blue,
    urlcolor=blue,
    colorlinks=true,
    pdfauthor={A.A. Golovanov, I.Yu. Kostyukov, J. Thomas, A. Pukhov},
    pdftitle={Analytic model for electromagnetic fields in the bubble regime of plasma wakefield in non-uniform plasmas}
}
\newcommand{\ion}{\textup{i}}
\newcommand{\electron}{\textup{e}}

\newcommand{\plasm}{\textup{p}}

\newcommand{\rb}{r_\mathrm{b}}
\newcommand{\bunch}{\textup{d}}
\newcommand{\channel}{\textup{c}}

\begin{document}
\title{Analytic model for electromagnetic fields in the bubble regime of plasma wakefield in non-uniform plasmas}
\author{A.\,A. Golovanov}
\author{I.\,Yu. Kostyukov}
\affiliation{Institute of Applied Physics RAS, 603950 Nizhny Novgorod, Russia}
\author{J. Thomas}
\author{A. Pukhov}
\affiliation{Institut für Theoretische Physik I, Heinrich-Heine-Universität Düsseldorf, Düsseldorf D-40225, Germany}

\begin{abstract}
    Based on a model of plasma wakefield in the strongly nonlinear (bubble) regime, we develop a lowest-order perturbation theory for the components of electromagnetic fields inside and outside the bubble using the assumption of small thickness of the electron sheath on the boundary of the bubble.
    Unlike previous models, we derive simple explicit expressions for the components of electromagnetic fields not only in the vicinity of the center of the bubble, but in the whole volume of the bubble (including areas of driving or accelerated bunches) as well as outside it.
    Moreover, we apply the results to the case of radially non-uniform plasma and, in particular, to plasma with a hollow channel.
    The obtained results are verified with 3D particle-in-cell (PIC) simulations which show good correspondence to our model.
\end{abstract}
\maketitle

\section*{Introduction}

Currently, plasma acceleration methods are considered promising for obtaining high-energy electron bunches. \cite{Esarey2009RevModPhys, Kostyukov2015UFN}
Their main idea is the use of a driver (a relativistic electron bunch \cite{Rosenzweig1988experimental} or an intense laser pulse\cite{Tajima1979laser}) for excitation of plasma wakefield whose longitudinal electric field can be used to accelerate charged particles.
The achievable accelerating gradients can be of orders of magnitude higher than those in conventional accelerators.
In experiments with the laser-wakefield acceleration (LWFA), electrons with the energy of \SI{4.2}{\GeV} have been obtained at the acceleration length of \SI{9}{\cm}.\cite{Leemans2014PRL}
For plasma-wakefield acceleration (PWFA), in which a relativistic bunch is used as the driver, the possibility of energy doubling from \SI{42}{\GeV} to \SI{85}{\GeV} has been experimentally demonstrated.\cite{Blumenfeld2007Nature}

Particularly interesting is the so-called ``bubble'' or ``blow-out'' regime of plasma wakefield in which plasma electrons are almost completely expelled by the driver, leading to the formation of a spherical cavity (or a ``bubble'') behind the driver.\cite{Pukhov2002Bubble}
On the boundary of this cavity, a thin electron sheath composed of the expelled electrons is formed.
The densities and currents in this sheath screen the surrounding plasma from the wakefield, so its structure is important for processes happening on the boundary of the plasma bubble and outside it.
However, no self-consistent theory of the bubble regime which can predict the full structure of the wakefield in the bubble regime based on the properties of the driver exists.

Due to its complex nonlinear nature, the bubble regime is commonly studied numerically with the use of the particle-in-cell (PIC) method. \cite{Pukhov2016PIC}
However, analytic description of this regime is also of substantial interest.
Different theoretical approaches have been used to make significant advancement in the understanding of this regime, which have led to the creation of phenomenological models of the bubble,\cite{Kostyukov_2004_PoP_11_115256, Kostyukov_2009_PRL_103_175003} the development of a similarity theory \cite{Gordienko_2005_PoP_12_043109} and an analytic model of the bubble's boundary.\cite{Lu_2006_PoP_13_056709} 
Following the discovery of the advantages of plasmas with hollow channels in the bubble regime,\cite{Pukhov2014Channel} this analytic model has recently been generalized to describe plasmas with non-uniform transverse density profile.\cite{Golovanov_2016_QE_46_295, Thomas_2016_PoP_23_053108}

The aforementioned theories focus mostly on the shape of the bubble and on the dynamics of relativistic accelerated particles, while paying less attention to the structure of electromagnetic field components inside the bubble and at its boundary.
Knowledge of this structure can be of significant interest for the processes of particle injection and self-injection into the wakefield,\cite{Kostyukov_2009_PRL_103_175003, Yi_2013_PoP_20_013108} when the particles under consideration are not ultra-relativistic.
In principle, the approach which can be used to calculate the spatial distributions of the fields is presented by \citet{Yi_2013_PoP_20_013108}
However, this approach involves extensive numerical calculation of derivatives and integrals.
In the current paper, we make an additional step and develop a perturbation theory based on the assumption that the ratio between the thickness of the electron sheath at the boundary of the bubble and the size of the bubble is small.
This assumption allows to significantly simplify all expressions in the lowest order with respect to this ratio.
The novelty of our work compared to the model by \citet{Yi_2013_PoP_20_013108} lies in the following three points: \emph{
    (i) we obtain simple explicit expressions for the components of the electromagnetic field both inside and outside the bubble; (ii) we do not make an \emph{a priori} assumption about the shape of the electron sheath and consider the most general case;
    (iii) our results are applied to the case of radially non-uniform plasmas.
}

In section \ref{sec:general}, we write out basic equations for the potentials and electromagnetic fields in any kind of plasma wakefield.
In section \ref{sec:bubble}, we introduce a model for the bubble regime which assumes general structure of the electron sheath on the bubble's boundary.
In section \ref{sec:thinSheath}, we make an assumption that the electron sheath is thin compared to the size of the bubble.
This allows us to obtain a simpler equation for the boundary of the bubble.
In section \ref{sec:fields}, based on the aforementioned assumption, we develop a perturbation theory for the components of the electromagnetic field inside and outside the bubble and find the distributions of all these components.
All derivations are accompanied and verified by PIC simulations for plasma with a hollow channel which show excellent agreement with our model.
    
\section{General equations}
\label{sec:general}
We consider boundless plasma in which a laser pulse or an electron driver propagates along the axis $z$ and excites plasma wakefield in it.
The plasma is assumed to be non-uniform perpendicular to the $z$-axis.
Under the assumption of axial symmetry, the electron density $n_0$ depends only on the coordinate $r$ in cylindrical geometry.
For simplicity, we use dimensionless units in which all charges are normalized to $e$, densities to $n_\plasm$, time to $\omega_\plasm^{-1}$, coordinates to $k_\plasm^{-1} = c/\omega_\plasm$, momenta and energies to $mc$ and $mc^2$, respectively, and electric and magnetic fields to $mc\omega_\plasm/e$.
Here, $e > 0$ is the elementary charge, $m$ is the electron mass, $c$ is the speed of light in vacuum, $n_\plasm$ is typical plasma density (for example, for the case of a plasma with a hollow channel it may be the density outside the channel), $\omega_\plasm = (4\pi e^2 n_\plasm/m)^{1/2}$ is a typical electron plasma frequency.

Plasma wakefield is propagating with the velocity determined by the velocity of the driver and is close to the speed of light (equal to $1$ in dimensionless units).
As the shape of the wakefield changes slowly during its propagation through plasma, we use the quasi-stationary approximation and assume that electromagnetic fields in the wakefield depend on the longitudinal coordinate $z$ and time $t$ through their combination $\xi = t - z$. 
Plasma fields are described using the vector potential $\vb{A}$ and the wakefield potential $\Psi = \varphi - A_z$, where $\varphi$ is the scalar potential.
The electromagnetic fields can be easily retrieved from the potentials,
\begin{align}
    &B_\phi = -\pdv{A_r}{\xi} - \pdv{A_z}{r},\label{eq:magneticFieldThroughPotentials}\\
    &E_z = \pdv{\Psi}{\xi},\quad E_r = - \pdv{\Psi}{r} + B_\phi. \label{eq:electricFieldThroughPotentials}
\end{align}
All other components of the electromagnetic field are equal to zero due to axial symmetry.

We use the Lorenz gauge for the potentials
\begin{equation}
    \frac{1}{r}\pdv{}{r} (r A_r) = - \pdv{\Psi}{\xi},
\end{equation}
which allows us to express $A_r$ through the wakefield potential
\begin{equation}
    A_r = -\frac{1}{r} \int_0^r {\pdv{\Psi(r',\xi)}{\xi}r' \dd{r'}},
    \label{eq:ArThroughPsi}
\end{equation}
thus leaving only $\Psi$ and $A_z$ as independent potentials.
The combined Maxwell's equations for these potentials under the Lorenz gauge can be separated
\begin{align}
    &\frac{1}{r} \pdv{}{r} \left(r \pdv{A_z}{r}\right) = - J_z,\\
    &\frac{1}{r} \pdv{}{r} \left(r \pdv{\Psi}{r}\right) = J_z - \rho,
\end{align}
where $J_z$ is the longitudinal current density, $\rho$ is the charge density.
By integrating these equations we get
\begin{align}
    &\pdv{A_z}{r} = -\frac{1}{r} \int_0^r {J_z(\xi,r') r' \dd{r'}}, \label{eq:AzThroughJz}\\
    &\Psi = \Psi_0(\xi) + \int_0^r {\frac{\dd{r'}}{r'} \int_0^{r'} r'' S(\xi,r'') \dd{r''}}. \label{eq:PsiThroughS}
\end{align}
Here, we introduce $S(\xi,r) = J_z - \rho$.
The equation for $A_z$ is integrated only once because its radial derivative is sufficient to retrieve the fields in Eqs.~(\ref{eq:magneticFieldThroughPotentials}), (\ref{eq:electricFieldThroughPotentials}).

So, if we know the spatial distributions of $S$ and $J_z$ in the wakefield, we can easily find spatial distributions of the electromagnetic field components.
So far, all equations can be applied to any kind of wakefield.
Next, we introduce phenomenological models of these two sources for the bubble regime of wakefield. 

\section{Model of the bubble}
\label{sec:bubble}

Based on the properties of the bubble regime observed in the PIC simulations, we introduce a model of a bubble in which we assume that no plasma electrons are present inside the bubble, and there is a thin electron sheath on its boundary determined by a function $\rb(\xi)$.
Far outside the bubble, the plasma is assumed non-perturbed.
In this case, the source $S = J_z - \rho$ is modeled as\cite{Golovanov_2016_QE_46_295}
\begin{equation}
    S(\xi,r) =
    \begin{dcases}
        -\rho_\ion(r), & r < \rb(\xi),\\
        S_0(\xi) g\left(\frac{r - \rb(\xi)}{\Delta}\right),& r \geq \rb(\xi).
    \end{dcases}
    \label{eq:SModel}
\end{equation}
Inside the bubble, $S = J_z - \rho = -\rho_\ion$, where $\rho_\ion$ is the ion charge density.
For relativistic electron bunches in the bubble, the longitudinal velocity is close to $1$ (the dimensionless speed of light), thus $J_z \approx \rho$, so they do not contribute to $S$.
The function $g(X)$ describes the shape of the electron sheath at the bubble's boundary.
We assume that it is normalized, so that $M_0(0) = 1$ and $M_1(0) = 1$, where
\begin{equation}
    M_n (X) = {\int_X^\infty Y^n g(Y) \dd{Y}}
    \label{eq:momentaDefinition}
\end{equation}
are generalized moments of the function $g$.
The parameter $\Delta$ determines the typical width of the electron sheath.
In the previous works, exponential $g(X) = \exp(-X)$ \cite{Yi_2013_PoP_20_013108} and rectangular $g(X) = \theta(1 - X)$ \cite{Lu_2006_PoP_13_056709, Thomas_2016_PoP_23_053108} profiles have been used.

The function $S_0(\xi)$ is obtained using the condition $\lim_{r\to\infty} \Psi = 0$ and Eq.~\eqref{eq:PsiThroughS}
\begin{equation}
    S_0(\xi) = \frac{S_\ion(\rb(\xi))}{\Delta^2 (1 + \epsilon^{-1}(\rb))},
    \label{eq:S0}
\end{equation}
where 
\begin{equation}
    S_\ion(r) = \int_0^r {\rho_\ion(r') r' \dd{r'}}, \quad \epsilon(\rb) = \Delta / \rb.
    \label{eq:Sion}
\end{equation}
The same answer for $S_0(\xi)$ can be obtained from the continuity equation and the assumption that $S_0(\xi) = 0$ when $\rb(\xi) = 0$.\cite{Thomas_2016_PoP_23_053108}

A similar model is used for the longitudinal current $J_z(\xi,r)$.
Inside the bubble, $J_z = J_\electron \approx \rho_\electron$ is the current created by relativistic electron bunches (either driver or witness bunches).
Thus, 
\begin{equation}
    J_z(\xi, r) =
    \begin{dcases}
        \rho_\electron(\xi,r), & r < \rb(\xi),\\
        J_0(\xi) g_J\left(\frac{r - \rb(\xi)}{\Delta_J}\right), & r \geq \rb(\xi).
    \end{dcases}
    \label{eq:jModel}
\end{equation}
As shown by \citet{Yi_2013_PoP_20_013108}, a different typical width $\Delta_J \neq \Delta$ is required to describe the longitudinal current distribution.
A different from $g(X)$ shape function $g_J(X)$ is also introduced for the purpose of generality.
This function is normalized in the same way as $g(X)$ with $M_{J,0} (0) = M_{J,0}(0) = 1$, where $M_{J,n}(X)$ are introduced similarly to Eq.~\eqref{eq:momentaDefinition}.
The value of $J_0(\xi)$ can be obtained from the condition $\lim_{r\to\infty} r B_\phi = 0$, which will be demonstrated in Sec.~\ref{sec:fields}.

So, for any specific shape of the bubble $\rb(\xi)$, the models \eqref{eq:SModel}, \eqref{eq:jModel} allow us to calculate all of the field distributions.
As shown by \citet{Golovanov_2016_QE_46_295}, model \eqref{eq:SModel} and electron motion equations lead to the equation for the bubble's boundary
\begin{equation}
    A(\rb) \rb'' + B(\rb) \rb'^2 + C(\rb) = \frac{\lambda(\xi)}{\rb},
    \label{eq:BubbleEqABC}
\end{equation}
where $\rb''= \dv*[2]{\rb(\xi)}{\xi}$, $\rb' = \dv*{\rb(\xi)}{\xi}$, 
\begin{equation}
    \lambda(\xi) = - \int_0^{\rb(\xi)} {\rho_\electron(\xi, r') r' \dd{r'}}
    \label{eq:Lambda}
\end{equation}
is the source term created by relativistic electron bunches inside the bubble,
and the coefficients $A$, $B$, $C$ are determined by the shape of the electron sheath $g(X)$ and its width $\Delta$.
Despite having no laser-related terms, this equation can be applied for the case of a laser driver in regions where the laser pulse is not present.
It is also possible to include the influence of the laser field on the boundary of the bubble into this equation, which is not considered in the current paper; for uniform plasma it is done in Vieira \emph{et~al.}\cite{Vieira2016BubbleSummary}

If we take $\rho_\ion(r) = 1$, $g(X) = g_J(X) = \exp(-X)$, our theory up to this point is completely equivalent to the model used by \citet{Yi_2013_PoP_20_013108}
However, despite the fact that all previous equations give an analytical model for the fields inside and outside the bubble, the expressions for the field distributions are too complex, so they can be obtained only numerically.
In the subsequent sections, we show that it is possible to obtain significantly simpler expressions by making an assumption about the width of the electron sheath.

\section{Thin sheath approximation}
\label{sec:thinSheath}

As shown in the previous works,\cite{Lu_2006_PoP_13_056709, Thomas_2016_PoP_23_053108, Golovanov_2016_QE_46_295} Eq.~\eqref{eq:BubbleEqABC} for the bubble's boundary can be significantly simplified by assuming that the width of the electron sheath on the boundary is significantly smaller than the size of the bubble, i.\,e. $\Delta \ll \rb$.
If we also assume that $\Delta \gtrsim \rb / S_\ion(\rb)$, which corresponds to a sufficiently large bubble, Eq.~\eqref{eq:BubbleEqABC} transforms into
\begin{equation}
    S_\ion(\rb) \rb \rb'' + \rho_\ion(\rb) \rb^2 \rb'^2 + S_\ion(\rb) = 2 \lambda(\xi),
    \label{eq:BubbleEq}
\end{equation}
where $S_\ion(r)$ is given by Eq.~\eqref{eq:Sion}.
The coefficients of this equation are determined solely by the plasma profile $\rho_\ion$, while the source on the right-hand side reflects the influence of relativistic electron bunches inside the bubble.

\begin{figure}
	\centering
    \includegraphics[]{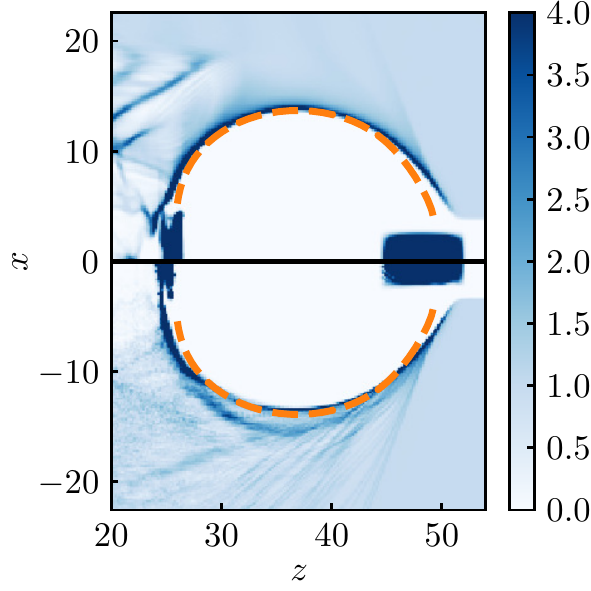}
	\caption{Electron density distribution in a bubble excited by a relativistic electron bunch in plasma with a hollow channel in PIC codes Smilei (top) and Quill (bottom).
	The analytical border of the bubble calculated using Eq.~\eqref{eq:BubbleEq} is shown with the dashed lines.
    All lengths are normalized to $k_\plasm^{-1} = \lambda_\plasm/2\pi$.}
    \label{fig:bubble}
\end{figure}

The applicability of this model has already been verified by PIC simulations.\cite{Thomas_2016_PoP_23_053108, Golovanov_2016_PoP_23_093114}
To demonstrate this again and to have numerical results for further analysis, we have carried out simulations of a wakefield generated by a relativistic electron bunch using the open-source 3D PIC code Smilei. \cite{Smilei, Derouillat2017smilei}
This code operates in the same dimensionless values as our theory, so all initial conditions were set dimensionless.
In these simulations, we used a driver (an electron bunch) with parabolic density profile, transverse size $\sigma_{\bunch,r} = 1.6\pi$, longitudinal size $\sigma_{\bunch,z} = 2.4\pi$, charge of \SI{6.4}{\nano\coulomb}, and electrons' energy of \SI{2}{\GeV}.
High electron energy was chosen to prevent the evolution of the driver.
The surrounding plasma had a hollow channel of radius $r_\channel=1.2\pi$, corresponding to the electron density of $n_0(r) = \theta(r-r_\channel)$, where $\theta(X)$ is the Heaviside step-function.
For comparison, similar simulations have been performed using the PIC code Quill developed in our group. \cite{Nerush2010Quill}
The comparison of the resulting bubble with the model given by Eq.~\eqref{eq:BubbleEq} is shown in Fig.~\ref{fig:bubble}.
As expected, this model produces overall good results which slightly diverge from the actual shape of the bubble when $\rb$ becomes close to $r_\channel$ and thus approximations used in our model are invalid.
The results of the simulations are generally similar in the two codes we have used, so all following figures will be made using only the results from the simulations with Smilei.

Interestingly enough, unlike Eq.~\eqref{eq:BubbleEqABC}, Eq.~\eqref{eq:BubbleEq} holds no dependence on the width of the electron sheath $\Delta$ or its shape $g(X)$ whatsoever.
Encouraged by this fact, we hope that a similar level of simplification can be achieved for the components of the electromagnetic field.
So, in all further calculations, we assume that $\Delta \ll \rb$, corresponding to $\epsilon(\rb) = \Delta / \rb \ll 1$.
We also make a similar assumption regarding the electron sheath current profile: $\Delta_J \ll \rb$.
Using these two assumptions and model $\eqref{eq:BubbleEq}$, we rigorously calculate distributions of fields both inside and outside the bubble in the lowest possible order with respect to $\epsilon(\rb)$ and $\epsilon_J(\rb) = \Delta_J / \rb$.

\section{Fields distributions}
\label{sec:fields}

\subsection{Wakefield potential}
\label{sec:wakefield}

In order to calculate the fields, we first need to obtain the distribution of the wakefield potential $\Psi$ given by Eq.~\eqref{eq:PsiThroughS}.
Equations \eqref{eq:PsiThroughS}, \eqref{eq:SModel}, \eqref{eq:S0} together with the condition $\lim_{r\to\infty} \Psi = 0$ give us 
\begin{multline}
    \Psi_0(\xi) = \int_0^{\rb} {\frac{S_\ion(r')}{r'} \dd{r'}} + \\
    + S_\ion(\rb) \int_0^\infty \frac{\epsilon \dd{Y}}{1 + \epsilon Y} \frac{M_0(Y) + \epsilon M_1(Y)}{1 + \epsilon}.
\end{multline}
The second integral is equal to $\epsilon \int_0^\infty {M_0(Y) \dd{Y}} =  \epsilon M_1(0) = \epsilon$ in the first order with respect to $\epsilon$.
Thus, wakefield potential \eqref{eq:PsiThroughS} inside the bubble is
\begin{equation}
    \Psi(\xi,r) \approx \int_r^{\rb} {\frac{S_\ion(r')}{r'} \dd{r'}} + S_\ion(\rb) \epsilon(\rb).
    \label{eq:PsiInside}
\end{equation}
The second term can be neglected for $r < \rb$ but is required for $\Psi$ to stay continuous at $r = \rb$.

Outside the bubble ($r > \rb$), the solution to Eq.~\eqref{eq:PsiThroughS} in the first order in $\epsilon$ is 
\begin{equation}
    \Psi(\xi, r) \approx S_\ion(\rb) \epsilon(\rb) \int_{R(r,\rb)}^\infty {M_0(Y) \dd{Y}},
    \label{eq:PsiOutside}
\end{equation}
where $R(r,\rb) = (r - \rb) / \Delta$.

\subsection{Longitudinal electric field}

Expressions \eqref{eq:PsiInside}, \eqref{eq:PsiOutside} for the potential are sufficient to obtain the longitudinal field $E_z$ by using Eq.~\eqref{eq:electricFieldThroughPotentials}.
For the field inside the bubble, we differentiate only the first term in Eq.~\eqref{eq:PsiInside} and get:
\begin{equation}
    E_z \approx \frac{S_\ion(\rb)}{\rb} \rb',
    \label{eq:EzInside}
\end{equation}
which is a well-known formula\cite{Golovanov_2016_QE_46_295, Thomas_2016_PoP_23_053108} for the field inside the bubble.
This field does not depend on the radial coordinate $r$ because $\rb$ depends only on $\xi$.

Differentiating Eq.~\eqref{eq:PsiOutside} and leaving only the zeroth-order term in $\epsilon$ leads to
\begin{equation}
    E_z \approx \frac{S_\ion(\rb)}{\rb} \rb' M_0(R)
    \label{eq:EzOutside}
\end{equation}
for the longitudinal field component outside the bubble $(r > \rb)$.
It vanishes to zero according to $M_0(R)$.
For example, if $g(R) = \exp(-R)$, then $M_0(R) = \exp(-R) = \exp(-(r-\rb)/\Delta)$, so the electric field $E_z$ shows exponential decay with the typical width $\Delta$.

\begin{figure}
    \centering
    \includegraphics[]{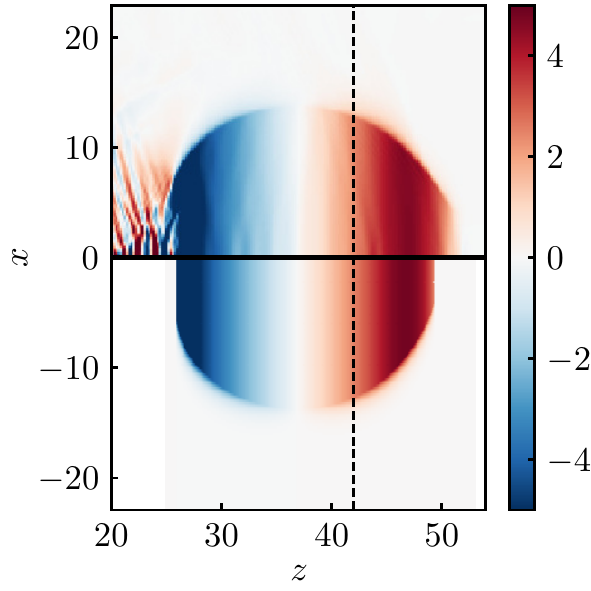}
    \caption{
        Numerical (top) and analytical (bottom) longitudinal electric field $E_z$ distributions in the bubble shown in Fig.~\ref{fig:bubble}.
        The analytical distribution is calculated using Eqs.~\eqref{eq:EzInside}, \eqref{eq:EzOutside}.
        The dashed line corresponds to the slice shown in Fig.~\ref{fig:EzSlices}(b).
    }
    \label{fig:Ez}
\end{figure}

\begin{figure}
    \centering
    \includegraphics[]{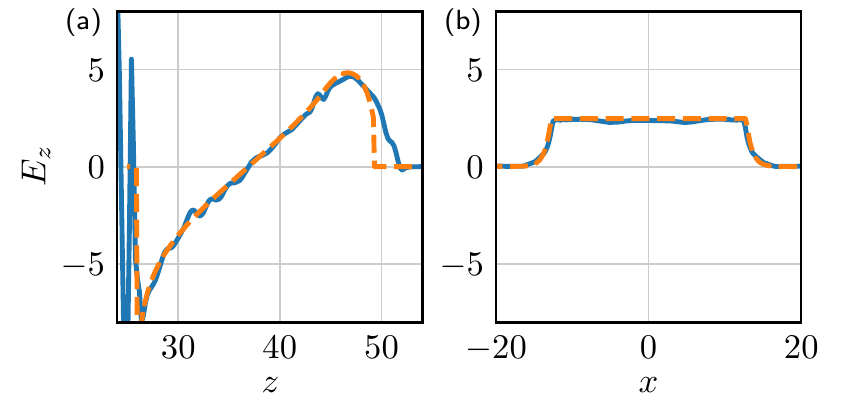}
    \caption{
        Longitudinal electric field $E_z$ distribution along (a) $x = 0$, (b) $z = 42$ corresponding to the distribution in Fig.~\ref{fig:Ez}.
        The solid and dashed lines correspond to the numerical and analytical results, respectively.
    }
    \label{fig:EzSlices}
\end{figure}

In order to compare the analytical results to the simulations, we need to determine the $g(X)$ function and the thickness $\Delta$ of the electron sheath.
Based on the results of the simulations corresponding to Fig.~\ref{fig:bubble}, we assume that $g(X) = \exp(-X)$.
In this case, $\Delta = 0.75$ gives the best fit to the simulations.
The comparison of the simulations results to analytical expressions \eqref{eq:EzInside}, \eqref{eq:EzOutside} is shown in Fig.~\ref{fig:Ez}.
Figure~\ref{fig:EzSlices} also shows longitudinal and transverse slices of the field distribution.
This comparison shows that our model adequately describes both the distribution of the electric field inside the bubble---which is indeed uniform in the transverse direction---and the way the electric field vanishes to zero outside the bubble.
As already mentioned, difference is seen only at the front edge of the bubble where our model is not applicable.

\subsection{Azimuthal magnetic field}

The magnetic field in our model has only the azimuthal component $B_\phi$.
According to Eqs.~\eqref{eq:magneticFieldThroughPotentials}, \eqref{eq:electricFieldThroughPotentials}, \eqref{eq:ArThroughPsi}, \eqref{eq:AzThroughJz}, it can be obtained by solving an equation
\begin{equation}
	rB_\phi(\xi, r) = \rb B_\phi(\xi, \rb) + \int_{\rb}^r {\left(J_z + \pdv{E_z}{\xi}\right) r' \dd{r'}}.
	\label{eq:BphiEq}
\end{equation}
By setting $r = 0$, we can get
\begin{equation}
	\rb B_\phi(\xi, \rb) = \int_0^{\rb} {\left(J_z + \pdv{E_z}{\xi}\right) r' \dd{r'}},
\end{equation}
which can be easily integrated.
Inside the bubble (for $r < \rb$), $J_z \approx \rho_\electron$, thus integrating the first term gives us $-\lambda(\xi)$ by its definition \eqref{eq:Lambda}.
As the longitudinal electric field inside the bubble, according to Eq.~\eqref{eq:EzInside}, does not depend on $r$, integration of the second term is reduced to multiplication by $\rb^2 / 2$.
Thus,
\begin{multline}
	\rb B_\phi(\xi, \rb) \approx -\lambda(\xi) + \\ + \frac{\rb^2}{2}\left[\rho_\ion(\rb) \rb'^2 - \frac{S_\ion(\rb)}{\rb^2} \rb'^2 +  \frac{S_\ion}{\rb}\rb'' \right] = \\ = -\frac{S_\ion(\rb)}{2}\left(1 + \rb'^2\right).
\end{multline}
To simplify the results, we have also used Eq.~\eqref{eq:BubbleEq}.

Using a similar approach, it is straightforward to calculate the magnetic field for $r < \rb$.
The result is
\begin{equation}
	B_\phi(\xi,r) \approx - r \left[\frac{S_\ion(\rb)}{2\rb^2} (1 + \rb'^2) -\frac{\lambda(\xi)}{\rb^2}\right] - \frac{\tilde\lambda(\xi,r)}{r},
	\label{eq:BphiInside}
\end{equation}
where 
\begin{equation}
	\tilde\lambda(\xi, r) = -\int_0^r {\rho_\electron(\xi, r') r' \dd{r'}}.
\end{equation}
Obviously, $\tilde\lambda(\xi, r=\rb) = \lambda(\xi)$.
So, for the longitudinal coordinates $\xi$ not located outside electron bunches ($\tilde\lambda = 0$), the azimuthal component of the magnetic fields grows linearly with $r$.
This linear behavior does not depend on the radial plasma profile.
Also, the sign of $B_\phi$ is always negative.

In order to calculate $B_\phi$ outside the bubble ($r > \rb$), we need to use model \eqref{eq:jModel} for $J_z$ and Eq.~\eqref{eq:EzOutside} for $E_z$ and substitute them into Eq.~\eqref{eq:BphiEq}.
The integration results in
\begin{align}
	&\int_{\rb}^r J_z r' \dd{r'} \approx J_0(\xi) \rb \Delta_J (1 - M_{J,0}(\alpha R)),\\
	&\int_{\rb}^r \pdv{E_z}{\xi} r' \dd{r'} \approx S_\ion(\rb) \rb'^2 (1 - M_0(R)),
\end{align}
where $\alpha = \Delta / \Delta_J$.
The value of $J_0(\xi)$ can be found from the condition $\lim_{r\to\infty} r B_\phi = 0$ and is
\begin{equation}
	J_0(\xi) = \frac{S_\ion(\rb)}{2\rb\Delta_J}\left(1 - \rb'^2\right).
	\label{eq:J0}
\end{equation}
Finally,
\begin{equation}
	B_\phi \approx \frac{S_\ion(\rb)}{2 \rb}\left[M_{J,0}(\alpha R)\left(\rb'^2 - 1\right) -2 M_0(R)\rb'^2\right].
	\label{eq:BphiOutside}
\end{equation}
This equation describes how $B_\phi$ converges to zero for $r > \rb$, and together with Eq.~\eqref{eq:BphiInside} they give the complete spatial distribution of the azimuthal magnetic field.
One more interesting result following from Eq.~\eqref{eq:J0} is that the longitudinal current $J_z$ changes its sign at the points $\xi$ where $\abs{\rb'(\xi)} = 1$, which has been observed in the simulations by \citet{Yi_2013_PoP_20_013108}

\begin{figure}
    \centering
    \includegraphics[]{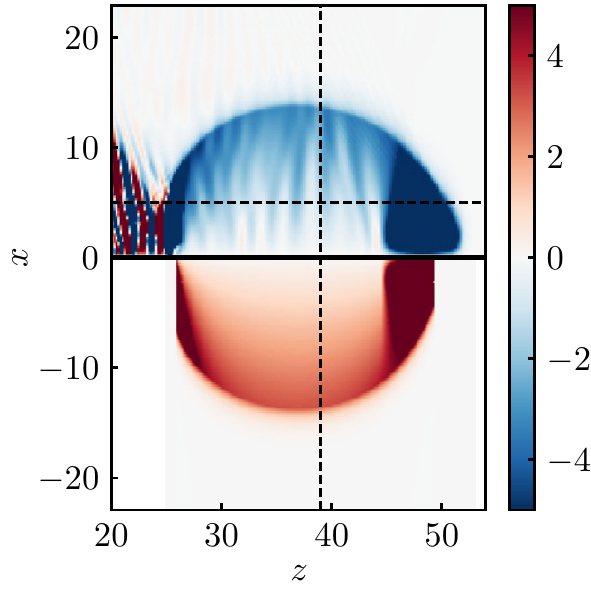}
    \caption{
        Numerical (top) and analytical (bottom) azimuthal magnetic field $B_y$ distributions in the bubble shown in Fig.~\ref{fig:bubble}.
        The analytical distribution is calculated using Eqs.~\eqref{eq:BphiInside}, \eqref{eq:BphiOutside}.
        The dashed lines correspond to slices shown in Fig.~\ref{fig:BphiSlices}.
    }
    \label{fig:Bphi}
\end{figure}

\begin{figure}
    \centering
    \includegraphics[]{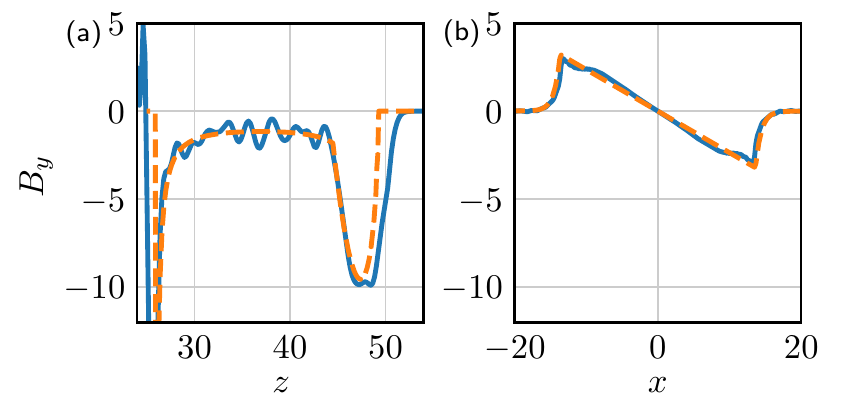}
    \caption{
        Azimuthal magnetic field distribution along (a) $x = 5$, (b) $z = 39$ corresponding to the distribution in Fig.~\ref{fig:Bphi}.
        The solid and dashed lines correspond to the numerical and analytical results, respectively.
    }
    \label{fig:BphiSlices}
\end{figure}

In order to perform comparison to the simulations, we have assumed $g_J(X) = \exp(-X)$ and set $\Delta_J = 0.81$ to provide the best fit to the results of the simulations.
The comparison is shown in Figs.~\ref{fig:Bphi} and \ref{fig:BphiSlices}.
It again shows the validity of our approximate theory for the description of the fields both inside and outside the bubble.
The longitudinal perturbations of the magnetic field in the simulations are not physical but numerical.
There are two numerical sources of these perturbations: numerical Cherenkov radiation from the relativistic driver\cite{godfrey1974numerical} and reflections from the simulation box boundaries.
The box used in simulations is large enough ($18\pi$ in all dimensions) to mitigate the reflections.
However, numerical Cherenkov radiation of ultrarelativistic electrons cannot be mitigated by changing the simulation parameters; methods for its suppression are not available yet in the Smilei 3D PIC code.

\subsection{Radial electric field}
After we have calculated the azimuthal component of the magnetic field, it is a very simple task to calculate the radial electric field as, according to Eq.~\eqref{eq:electricFieldThroughPotentials}, it is obtained from $B_\phi$ by adding $-\pdv*{\Psi}{r}$, and $\Psi$ has already been calculated in Sec.~\ref{sec:wakefield}.
Thus, for $r < \rb$, we obtain
\begin{equation}
	E_r(\xi,r) = B_\phi(\xi, r) + \frac{S_\ion(r)}{r},
	\label{eq:ErInside}
\end{equation}
where $B_\phi$ is given by Eq.~\eqref{eq:BphiInside}, and for $r > \rb$, we have
\begin{equation}
	E_r \approx \frac{S_\ion(\rb)}{2\rb} \left[M_{J,0}(\alpha R) - 2 M_0(R)\right] \left(\rb'^2 - 1\right).
	\label{eq:ErOutside}
\end{equation}
Inside the bubble, $S_\ion(r)/r$ and $B_\phi$ have different signs.
For uniform plasma, $S_\ion(r)/r = r/2$ grows linearly with $r$ and is typically larger than $B_\phi$, thus $E_r$ has a different sign than $B_\phi$ and also shows linear dependence on $r$, which is well-known from previous models. \cite{Kostyukov_2004_PoP_11_115256}
However, for plasmas with a hollow channel, $S_\ion(r) = 0$ inside the channel, and therefore $E_r$ can have different signs inside and outside the channel.
This property of the radial electric field in a hollow channel has been observed by \citet{Pukhov2014Channel}
Also, the radial component of the electric field, like $J_z$, turns to zero at the boundary of the bubble when $\abs{\rb'} = 1$.

\begin{figure}
    \centering
    \includegraphics[]{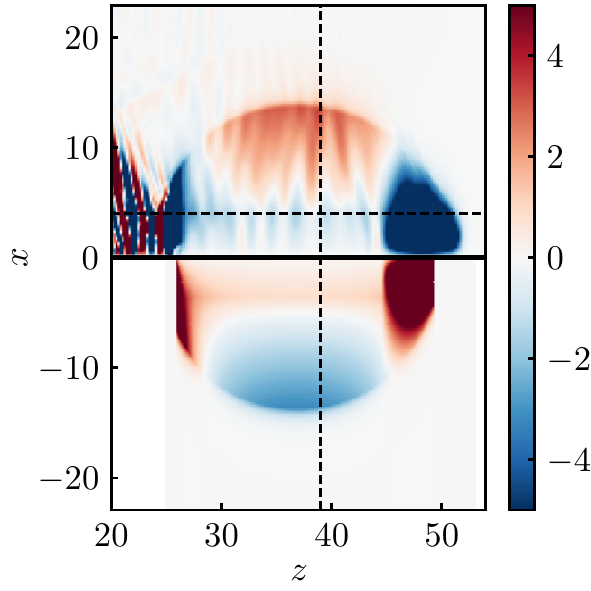}
    \caption{
        Numerical (top) and analytical (bottom) radial electric field $E_x$ distributions in the bubble shown in Fig.~\ref{fig:bubble}.
        The analytical distribution is calculated using Eqs.~\eqref{eq:ErInside}, \eqref{eq:ErOutside}.
        The dashed lines correspond to the slices shown in Fig.~\ref{fig:ErSlices}.
    }
    \label{fig:Er}
\end{figure}

\begin{figure}
    \centering
    \includegraphics[]{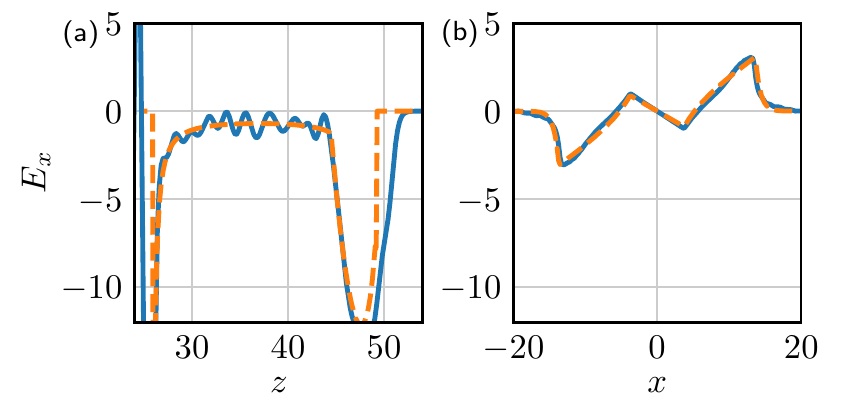}
    \caption{
        Radial electric field distribution along (a) $x = 4$, (b) $z = 39$ corresponding to the distribution in Fig.~\ref{fig:Er}.
        The solid and dashed lines correspond to the numerical and analytical results, respectively.
    }
    \label{fig:ErSlices}
\end{figure}

We observe this behavior in the PIC simulations shown in Figs.~\ref{fig:Er}, \ref{fig:ErSlices}.
The presence of the hollow channel in plasma leads to the change in the sign of $E_r$ outside it.
Again, the longitudinal perturbations of the field are the result of the numerical Cherenkov radiation and reflections from the walls.

\subsection{Radial force}

The Lorentz force acting on accelerated particles in the bubble is important to understand their dynamics.
General distributions of $E_z$, $E_r$, and $B_\phi$ are sufficient to find the force acting on any particle.
However, there is a special case of relativistic particles which move alongside the wakefield with $\vb{v} \approx c \vb{z}_0$.
This special case is typical for particles accelerated in the bubble and for the driver; for this case, the Lorentz force is significantly simplified. \cite{Kostyukov_2004_PoP_11_115256}
The longitudinal component of the Lorentz force acting on such an electron is determined only by the longitudinal field distribution, $F_z \approx - E_z = -\pdv*{\Psi}{\xi}$, while the transverse force 
\begin{equation}
    F_r = - E_r - (\vb{v} \times \vb{B})_r \approx -E_r + B_\phi = \pdv{\Psi}{r}.
\end{equation}
These forces depend only on the distribution of the wakefield potential $\Psi$.

As we have already calculated the distributions of $E_r$ and $B_\phi$, the distribution of $F_r$ is calculated simply by taking the difference between them.
The force is
\begin{equation}
    F_r(\xi,r) = 
    \begin{dcases}
        -\frac{S_\ion(r)}{r}, & r < \rb(\xi),\\
        -\frac{S_\ion(\rb)}{\rb} M_0(R), & r \geq \rb(\xi).
    \end{dcases}
    \label{eq:Fr}
\end{equation}
This force is always focusing for electrons (and defocusing for positrons) and turns to zero inside a hollow channel where $S_\ion(r) = 0$.

\begin{figure}
    \centering
    \includegraphics[]{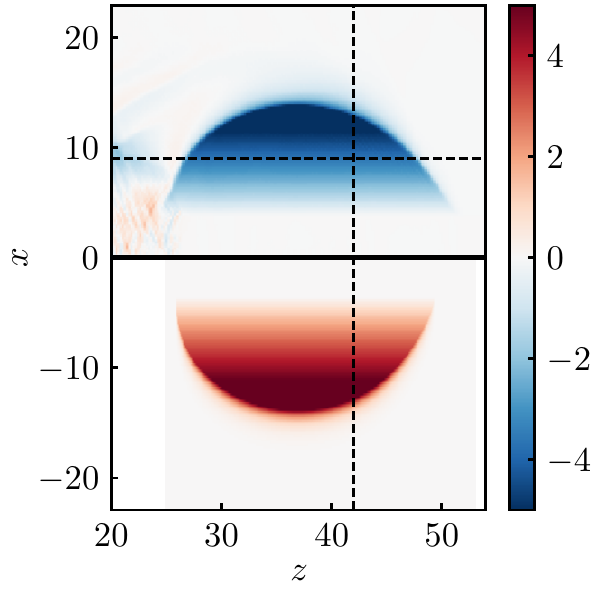}
    \caption{
        Numerical (top) and analytical (bottom) radial force $F_x = B_y - E_x$ distributions in the bubble shown in Fig.~\ref{fig:bubble}.
        The analytical distribution is calculated using Eq.~\eqref{eq:Fr}.
        The dashed lines correspond to the slices shown in Fig.~\ref{fig:FrSlices}.
    }
    \label{fig:Fr}
\end{figure}

\begin{figure}
    \centering
    \includegraphics[]{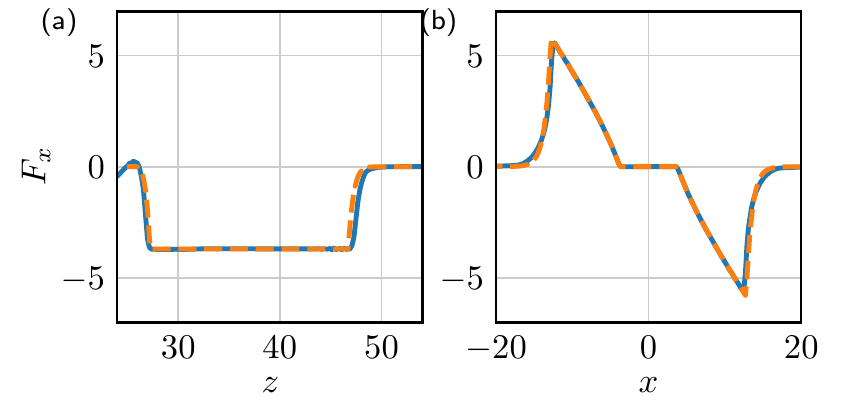}
    \caption{
        Radial force distribution along (a) $x = 9$, (b) $z = 42$ corresponding to the distribution in Fig.~\ref{fig:Fr}.
        The solid and dashed lines correspond to the numerical and analytical results, respectively.
    }
    \label{fig:FrSlices}
\end{figure}

The comparison of analytical and numerical distributions is shown in Figs.~\ref{fig:Fr} and \ref{fig:FrSlices}.
Unlike distributions of $E_r$ and $B_\phi$, the distribution of $F_r = B_\phi - E_r$ obtained from PIC simulations shows almost perfect correspondence to the one predicted by our model, which further confirms that the longitudinal perturbations visible in Figs.~\ref{fig:Bphi} and \ref{fig:Er} are transverse electromagnetic waves radiated by a driver via the mechanism of numerical Cherenkov radiation.

\section*{Conclusions}

In this paper, we have considered an analytical model of a bubble excited by a relativistic electron bunch or a laser pulse in transversely non-uniform plasma.
Special attention in our theory is paid to the electron sheath formed at the boundary of the bubble.
The distributions of densities and currents in this sheath is important for the properties of the electromagnetic field outside the bubble.
Unlike most previous theories, we use a more general model for this electron sheath.
Using the assumption that the electron sheath is thin compared to the size of the bubble, we have developed a lowest-order perturbation theory and obtained explicit expressions for all components of the electromagnetic field both inside and outside the bubble.
3D particle-in-cell simulations with two PIC codes (Smilei and Quill) confirm the validity of our model.
Noticeable difference between our model and the simulations is observed only in the small areas near the front and the back edges of the bubble, where our approximate theory is not applicable.
Hence, the obtained simple analytic expressions can be used to describe the Lorentz force acting on charged particles in the wakefield and therefore their motion.
The dynamics of charged particles can be of considerable interest for the processes of self-injection and external injection which play important role in determining the properties of the accelerated bunches.

Despite the fact that our theory is able to correctly predict the shape of the bubble and fields both inside and outside it, it still requires knowledge of several parameters from simulations, including the profile and the width of the electron sheath and the size of the bubble.
The possibility of creation of a self-consistent theory which is able to predict the full structure of the wakefield based on the properties of the driver is still an important task for further investigation.
One more important phenomenon our model does not consider are the so called bow waves \cite{Esirkepov2008BowWave, Luo2016waves} which are streams of electrons detaching from the bubble.
This phenomenon is visible in our simulations in Fig.~\ref{fig:bubble}, but it does not have a significant influence on the fields spatial distribution for the parameters used in the simulations.
However, the possibility of this phenomenon being more important under different set of parameters is a topic for future consideration.

\begin{acknowledgments}
    This work has been supported by the Russian Science Foundation through Grant No.\,16-12-10383.
\end{acknowledgments}

\section*{References}
\bibliographystyle{aipnum4-1}
\bibliography{BibliographyPop}

\begin{thebibliography}{24}%
\makeatletter
\providecommand \@ifxundefined [1]{%
 \@ifx{#1\undefined}
}%
\providecommand \@ifnum [1]{%
 \ifnum #1\expandafter \@firstoftwo
 \else \expandafter \@secondoftwo
 \fi
}%
\providecommand \@ifx [1]{%
 \ifx #1\expandafter \@firstoftwo
 \else \expandafter \@secondoftwo
 \fi
}%
\providecommand \natexlab [1]{#1}%
\providecommand \enquote  [1]{``#1''}%
\providecommand \bibnamefont  [1]{#1}%
\providecommand \bibfnamefont [1]{#1}%
\providecommand \citenamefont [1]{#1}%
\providecommand \href@noop [0]{\@secondoftwo}%
\providecommand \href [0]{\begingroup \@sanitize@url \@href}%
\providecommand \@href[1]{\@@startlink{#1}\@@href}%
\providecommand \@@href[1]{\endgroup#1\@@endlink}%
\providecommand \@sanitize@url [0]{\catcode `\\12\catcode `\$12\catcode
  `\&12\catcode `\#12\catcode `\^12\catcode `\_12\catcode `\%12\relax}%
\providecommand \@@startlink[1]{}%
\providecommand \@@endlink[0]{}%
\providecommand \url  [0]{\begingroup\@sanitize@url \@url }%
\providecommand \@url [1]{\endgroup\@href {#1}{\urlprefix }}%
\providecommand \urlprefix  [0]{URL }%
\providecommand \Eprint [0]{\href }%
\providecommand \doibase [0]{http://dx.doi.org/}%
\providecommand \selectlanguage [0]{\@gobble}%
\providecommand \bibinfo  [0]{\@secondoftwo}%
\providecommand \bibfield  [0]{\@secondoftwo}%
\providecommand \translation [1]{[#1]}%
\providecommand \BibitemOpen [0]{}%
\providecommand \bibitemStop [0]{}%
\providecommand \bibitemNoStop [0]{.\EOS\space}%
\providecommand \EOS [0]{\spacefactor3000\relax}%
\providecommand \BibitemShut  [1]{\csname bibitem#1\endcsname}%
\let\auto@bib@innerbib\@empty
\bibitem [{\citenamefont {Esarey}, \citenamefont {Schroeder},\ and\
  \citenamefont {Leemans}(2009)}]{Esarey2009RevModPhys}%
  \BibitemOpen
  \bibfield  {author} {\bibinfo {author} {\bibfnamefont {E.}~\bibnamefont
  {Esarey}}, \bibinfo {author} {\bibfnamefont {C.~B.}\ \bibnamefont
  {Schroeder}}, \ and\ \bibinfo {author} {\bibfnamefont {W.~P.}\ \bibnamefont
  {Leemans}},\ }\href {\doibase 10.1103/RevModPhys.81.1229} {\bibfield
  {journal} {\bibinfo  {journal} {Rev. Mod. Phys.}\ }\textbf {\bibinfo {volume}
  {81}},\ \bibinfo {pages} {1229} (\bibinfo {year} {2009})}\BibitemShut
  {NoStop}%
\bibitem [{\citenamefont {Kostyukov}\ and\ \citenamefont
  {Pukhov}(2015)}]{Kostyukov2015UFN}%
  \BibitemOpen
  \bibfield  {author} {\bibinfo {author} {\bibfnamefont {I.~{\relax Yu}.}\
  \bibnamefont {Kostyukov}}\ and\ \bibinfo {author} {\bibfnamefont {A.~M.}\
  \bibnamefont {Pukhov}},\ }\href {\doibase 10.3367/UFNe.0185.201501g.0089}
  {\bibfield  {journal} {\bibinfo  {journal} {Phys.-Usp.}\ }\textbf {\bibinfo
  {volume} {58}},\ \bibinfo {pages} {81} (\bibinfo {year} {2015})}\BibitemShut
  {NoStop}%
\bibitem [{\citenamefont {Rosenzweig}\ \emph {et~al.}(1988)\citenamefont
  {Rosenzweig}, \citenamefont {Cline}, \citenamefont {Cole}, \citenamefont
  {Figueroa}, \citenamefont {Gai}, \citenamefont {Konecny}, \citenamefont
  {Norem}, \citenamefont {Schoessow},\ and\ \citenamefont
  {Simpson}}]{Rosenzweig1988experimental}%
  \BibitemOpen
  \bibfield  {author} {\bibinfo {author} {\bibfnamefont {J.~B.}\ \bibnamefont
  {Rosenzweig}}, \bibinfo {author} {\bibfnamefont {D.~B.}\ \bibnamefont
  {Cline}}, \bibinfo {author} {\bibfnamefont {B.}~\bibnamefont {Cole}},
  \bibinfo {author} {\bibfnamefont {H.}~\bibnamefont {Figueroa}}, \bibinfo
  {author} {\bibfnamefont {W.}~\bibnamefont {Gai}}, \bibinfo {author}
  {\bibfnamefont {R.}~\bibnamefont {Konecny}}, \bibinfo {author} {\bibfnamefont
  {J.}~\bibnamefont {Norem}}, \bibinfo {author} {\bibfnamefont
  {P.}~\bibnamefont {Schoessow}}, \ and\ \bibinfo {author} {\bibfnamefont
  {J.}~\bibnamefont {Simpson}},\ }\href {\doibase 10.1103/PhysRevLett.61.98}
  {\bibfield  {journal} {\bibinfo  {journal} {Phys. Rev. Lett.}\ }\textbf
  {\bibinfo {volume} {61}},\ \bibinfo {pages} {98} (\bibinfo {year}
  {1988})}\BibitemShut {NoStop}%
\bibitem [{\citenamefont {Tajima}\ and\ \citenamefont
  {Dawson}(1979)}]{Tajima1979laser}%
  \BibitemOpen
  \bibfield  {author} {\bibinfo {author} {\bibfnamefont {T.}~\bibnamefont
  {Tajima}}\ and\ \bibinfo {author} {\bibfnamefont {J.~M.}\ \bibnamefont
  {Dawson}},\ }\href {\doibase 10.1103/PhysRevLett.43.267} {\bibfield
  {journal} {\bibinfo  {journal} {Phys. Rev. Lett.}\ }\textbf {\bibinfo
  {volume} {43}},\ \bibinfo {pages} {267} (\bibinfo {year} {1979})}\BibitemShut
  {NoStop}%
\bibitem [{\citenamefont {Leemans}\ \emph {et~al.}(2014)\citenamefont
  {Leemans}, \citenamefont {Gonsalves}, \citenamefont {Mao}, \citenamefont
  {Nakamura}, \citenamefont {Benedetti}, \citenamefont {Schroeder},
  \citenamefont {T{\'o}th}, \citenamefont {Daniels}, \citenamefont
  {Mittelberger}, \citenamefont {Bulanov}, \citenamefont {Vay}, \citenamefont
  {Geddes},\ and\ \citenamefont {Esarey}}]{Leemans2014PRL}%
  \BibitemOpen
  \bibfield  {author} {\bibinfo {author} {\bibfnamefont {W.~P.}\ \bibnamefont
  {Leemans}}, \bibinfo {author} {\bibfnamefont {A.~J.}\ \bibnamefont
  {Gonsalves}}, \bibinfo {author} {\bibfnamefont {H.-S.}\ \bibnamefont {Mao}},
  \bibinfo {author} {\bibfnamefont {K.}~\bibnamefont {Nakamura}}, \bibinfo
  {author} {\bibfnamefont {C.}~\bibnamefont {Benedetti}}, \bibinfo {author}
  {\bibfnamefont {C.~B.}\ \bibnamefont {Schroeder}}, \bibinfo {author}
  {\bibfnamefont {{\relax Cs}.}~\bibnamefont {T{\'o}th}}, \bibinfo {author}
  {\bibfnamefont {J.}~\bibnamefont {Daniels}}, \bibinfo {author} {\bibfnamefont
  {D.~E.}\ \bibnamefont {Mittelberger}}, \bibinfo {author} {\bibfnamefont
  {S.~S.}\ \bibnamefont {Bulanov}}, \bibinfo {author} {\bibfnamefont {J.-L.}\
  \bibnamefont {Vay}}, \bibinfo {author} {\bibfnamefont {C.~G.~R.}\
  \bibnamefont {Geddes}}, \ and\ \bibinfo {author} {\bibfnamefont
  {E.}~\bibnamefont {Esarey}},\ }\href {\doibase
  10.1103/PhysRevLett.113.245002} {\bibfield  {journal} {\bibinfo  {journal}
  {Phys. Rev. Lett.}\ }\textbf {\bibinfo {volume} {113}},\ \bibinfo {pages}
  {245002} (\bibinfo {year} {2014})}\BibitemShut {NoStop}%
\bibitem [{\citenamefont {Blumenfeld}\ \emph {et~al.}(2007)\citenamefont
  {Blumenfeld}, \citenamefont {Clayton}, \citenamefont {Decker}, \citenamefont
  {Hogan}, \citenamefont {Huang}, \citenamefont {Ischebeck}, \citenamefont
  {Iverson}, \citenamefont {Joshi}, \citenamefont {Katsouleas}, \citenamefont
  {Kirby}, \citenamefont {Lu}, \citenamefont {Marsh}, \citenamefont {Mori},
  \citenamefont {Muggli}, \citenamefont {Oz}, \citenamefont {Siemann},
  \citenamefont {Walz},\ and\ \citenamefont {Zhou}}]{Blumenfeld2007Nature}%
  \BibitemOpen
  \bibfield  {author} {\bibinfo {author} {\bibfnamefont {I.}~\bibnamefont
  {Blumenfeld}}, \bibinfo {author} {\bibfnamefont {C.~E.}\ \bibnamefont
  {Clayton}}, \bibinfo {author} {\bibfnamefont {F.-J.}\ \bibnamefont {Decker}},
  \bibinfo {author} {\bibfnamefont {M.~J.}\ \bibnamefont {Hogan}}, \bibinfo
  {author} {\bibfnamefont {C.}~\bibnamefont {Huang}}, \bibinfo {author}
  {\bibfnamefont {R.}~\bibnamefont {Ischebeck}}, \bibinfo {author}
  {\bibfnamefont {R.}~\bibnamefont {Iverson}}, \bibinfo {author} {\bibfnamefont
  {C.}~\bibnamefont {Joshi}}, \bibinfo {author} {\bibfnamefont
  {T.}~\bibnamefont {Katsouleas}}, \bibinfo {author} {\bibfnamefont
  {N.}~\bibnamefont {Kirby}}, \bibinfo {author} {\bibfnamefont
  {W.}~\bibnamefont {Lu}}, \bibinfo {author} {\bibfnamefont {K.~A.}\
  \bibnamefont {Marsh}}, \bibinfo {author} {\bibfnamefont {W.~B.}\ \bibnamefont
  {Mori}}, \bibinfo {author} {\bibfnamefont {P.}~\bibnamefont {Muggli}},
  \bibinfo {author} {\bibfnamefont {E.}~\bibnamefont {Oz}}, \bibinfo {author}
  {\bibfnamefont {R.~H.}\ \bibnamefont {Siemann}}, \bibinfo {author}
  {\bibfnamefont {D.}~\bibnamefont {Walz}}, \ and\ \bibinfo {author}
  {\bibfnamefont {M.}~\bibnamefont {Zhou}},\ }\href {\doibase
  10.1038/nature05538} {\bibfield  {journal} {\bibinfo  {journal} {Nature}\
  }\textbf {\bibinfo {volume} {445}},\ \bibinfo {pages} {741} (\bibinfo {year}
  {2007})}\BibitemShut {NoStop}%
\bibitem [{\citenamefont {Pukhov}\ and\ \citenamefont
  {{Meyer-ter-Vehn}}(2002)}]{Pukhov2002Bubble}%
  \BibitemOpen
  \bibfield  {author} {\bibinfo {author} {\bibfnamefont {A.}~\bibnamefont
  {Pukhov}}\ and\ \bibinfo {author} {\bibfnamefont {J.}~\bibnamefont
  {{Meyer-ter-Vehn}}},\ }\href {\doibase 10.1007/s003400200795} {\bibfield
  {journal} {\bibinfo  {journal} {Appl. Phys. B}\ }\textbf {\bibinfo {volume}
  {74}},\ \bibinfo {pages} {355} (\bibinfo {year} {2002})}\BibitemShut
  {NoStop}%
\bibitem [{\citenamefont {Pukhov}(2016)}]{Pukhov2016PIC}%
  \BibitemOpen
  \bibfield  {author} {\bibinfo {author} {\bibfnamefont {A.}~\bibnamefont
  {Pukhov}},\ }\href {\doibase 10.5170/CERN-2016-001.181} {\bibfield  {journal}
  {\bibinfo  {journal} {CERN Yellow Rep.}\ }\textbf {\bibinfo {volume} {1}},\
  \bibinfo {pages} {181} (\bibinfo {year} {2016})}\BibitemShut {NoStop}%
\bibitem [{\citenamefont {Kostyukov}, \citenamefont {Pukhov},\ and\
  \citenamefont {Kiselev}(2004)}]{Kostyukov_2004_PoP_11_115256}%
  \BibitemOpen
  \bibfield  {author} {\bibinfo {author} {\bibfnamefont {I.}~\bibnamefont
  {Kostyukov}}, \bibinfo {author} {\bibfnamefont {A.}~\bibnamefont {Pukhov}}, \
  and\ \bibinfo {author} {\bibfnamefont {S.}~\bibnamefont {Kiselev}},\ }\href
  {\doibase 10.1063/1.1799371} {\bibfield  {journal} {\bibinfo  {journal}
  {Phys. Plasmas}\ }\textbf {\bibinfo {volume} {11}},\ \bibinfo {pages} {5256}
  (\bibinfo {year} {2004})}\BibitemShut {NoStop}%
\bibitem [{\citenamefont {Kostyukov}\ \emph {et~al.}(2009)\citenamefont
  {Kostyukov}, \citenamefont {Nerush}, \citenamefont {Pukhov},\ and\
  \citenamefont {Seredov}}]{Kostyukov_2009_PRL_103_175003}%
  \BibitemOpen
  \bibfield  {author} {\bibinfo {author} {\bibfnamefont {I.}~\bibnamefont
  {Kostyukov}}, \bibinfo {author} {\bibfnamefont {E.}~\bibnamefont {Nerush}},
  \bibinfo {author} {\bibfnamefont {A.}~\bibnamefont {Pukhov}}, \ and\ \bibinfo
  {author} {\bibfnamefont {V.}~\bibnamefont {Seredov}},\ }\href {\doibase
  10.1103/PhysRevLett.103.175003} {\bibfield  {journal} {\bibinfo  {journal}
  {Phys. Rev. Lett.}\ }\textbf {\bibinfo {volume} {103}},\ \bibinfo {pages}
  {175003} (\bibinfo {year} {2009})}\BibitemShut {NoStop}%
\bibitem [{\citenamefont {Gordienko}\ and\ \citenamefont
  {Pukhov}(2005)}]{Gordienko_2005_PoP_12_043109}%
  \BibitemOpen
  \bibfield  {author} {\bibinfo {author} {\bibfnamefont {S.}~\bibnamefont
  {Gordienko}}\ and\ \bibinfo {author} {\bibfnamefont {A.}~\bibnamefont
  {Pukhov}},\ }\href {\doibase 10.1063/1.1884126} {\bibfield  {journal}
  {\bibinfo  {journal} {Phys. Plasmas}\ }\textbf {\bibinfo {volume} {12}},\
  \bibinfo {pages} {043109} (\bibinfo {year} {2005})}\BibitemShut {NoStop}%
\bibitem [{\citenamefont {Lu}\ \emph {et~al.}(2006)\citenamefont {Lu},
  \citenamefont {Huang}, \citenamefont {Zhou}, \citenamefont {Tzoufras},
  \citenamefont {Tsung}, \citenamefont {Mori},\ and\ \citenamefont
  {Katsouleas}}]{Lu_2006_PoP_13_056709}%
  \BibitemOpen
  \bibfield  {author} {\bibinfo {author} {\bibfnamefont {W.}~\bibnamefont
  {Lu}}, \bibinfo {author} {\bibfnamefont {C.}~\bibnamefont {Huang}}, \bibinfo
  {author} {\bibfnamefont {M.}~\bibnamefont {Zhou}}, \bibinfo {author}
  {\bibfnamefont {M.}~\bibnamefont {Tzoufras}}, \bibinfo {author}
  {\bibfnamefont {F.~S.}\ \bibnamefont {Tsung}}, \bibinfo {author}
  {\bibfnamefont {W.~B.}\ \bibnamefont {Mori}}, \ and\ \bibinfo {author}
  {\bibfnamefont {T.}~\bibnamefont {Katsouleas}},\ }\href {\doibase
  10.1063/1.2203364} {\bibfield  {journal} {\bibinfo  {journal} {Phys.
  Plasmas}\ }\textbf {\bibinfo {volume} {13}},\ \bibinfo {pages} {056709}
  (\bibinfo {year} {2006})}\BibitemShut {NoStop}%
\bibitem [{\citenamefont {Pukhov}\ \emph {et~al.}(2014)\citenamefont {Pukhov},
  \citenamefont {Jansen}, \citenamefont {Tueckmantel}, \citenamefont {Thomas},\
  and\ \citenamefont {Kostyukov}}]{Pukhov2014Channel}%
  \BibitemOpen
  \bibfield  {author} {\bibinfo {author} {\bibfnamefont {A.}~\bibnamefont
  {Pukhov}}, \bibinfo {author} {\bibfnamefont {O.}~\bibnamefont {Jansen}},
  \bibinfo {author} {\bibfnamefont {T.}~\bibnamefont {Tueckmantel}}, \bibinfo
  {author} {\bibfnamefont {J.}~\bibnamefont {Thomas}}, \ and\ \bibinfo {author}
  {\bibfnamefont {I.~{\relax Yu}.}\ \bibnamefont {Kostyukov}},\ }\href
  {\doibase 10.1103/PhysRevLett.113.245003} {\bibfield  {journal} {\bibinfo
  {journal} {Phys. Rev. Lett.}\ }\textbf {\bibinfo {volume} {113}},\ \bibinfo
  {pages} {245003} (\bibinfo {year} {2014})}\BibitemShut {NoStop}%
\bibitem [{\citenamefont {Golovanov}\ \emph
  {et~al.}(2016{\natexlab{a}})\citenamefont {Golovanov}, \citenamefont
  {Kostyukov}, \citenamefont {Pukhov},\ and\ \citenamefont
  {Thomas}}]{Golovanov_2016_QE_46_295}%
  \BibitemOpen
  \bibfield  {author} {\bibinfo {author} {\bibfnamefont {A.~A.}\ \bibnamefont
  {Golovanov}}, \bibinfo {author} {\bibfnamefont {I.~{\relax Yu}.}\
  \bibnamefont {Kostyukov}}, \bibinfo {author} {\bibfnamefont {A.~M.}\
  \bibnamefont {Pukhov}}, \ and\ \bibinfo {author} {\bibfnamefont
  {J.}~\bibnamefont {Thomas}},\ }\href {\doibase 10.1070/QEL16040} {\bibfield
  {journal} {\bibinfo  {journal} {Quantum Electron.}\ }\textbf {\bibinfo
  {volume} {46}},\ \bibinfo {pages} {295} (\bibinfo {year}
  {2016}{\natexlab{a}})}\BibitemShut {NoStop}%
\bibitem [{\citenamefont {Thomas}\ \emph {et~al.}(2016)\citenamefont {Thomas},
  \citenamefont {Kostyukov}, \citenamefont {Pronold}, \citenamefont
  {Golovanov},\ and\ \citenamefont {Pukhov}}]{Thomas_2016_PoP_23_053108}%
  \BibitemOpen
  \bibfield  {author} {\bibinfo {author} {\bibfnamefont {J.}~\bibnamefont
  {Thomas}}, \bibinfo {author} {\bibfnamefont {I.~{\relax Yu}.}\ \bibnamefont
  {Kostyukov}}, \bibinfo {author} {\bibfnamefont {J.}~\bibnamefont {Pronold}},
  \bibinfo {author} {\bibfnamefont {A.}~\bibnamefont {Golovanov}}, \ and\
  \bibinfo {author} {\bibfnamefont {A.}~\bibnamefont {Pukhov}},\ }\href
  {\doibase 10.1063/1.4948712} {\bibfield  {journal} {\bibinfo  {journal}
  {Phys. Plasmas}\ }\textbf {\bibinfo {volume} {23}},\ \bibinfo {pages}
  {053108} (\bibinfo {year} {2016})}\BibitemShut {NoStop}%
\bibitem [{\citenamefont {Yi}\ \emph {et~al.}(2013)\citenamefont {Yi},
  \citenamefont {Khudik}, \citenamefont {Siemon},\ and\ \citenamefont
  {Shvets}}]{Yi_2013_PoP_20_013108}%
  \BibitemOpen
  \bibfield  {author} {\bibinfo {author} {\bibfnamefont {S.~A.}\ \bibnamefont
  {Yi}}, \bibinfo {author} {\bibfnamefont {V.}~\bibnamefont {Khudik}}, \bibinfo
  {author} {\bibfnamefont {C.}~\bibnamefont {Siemon}}, \ and\ \bibinfo {author}
  {\bibfnamefont {G.}~\bibnamefont {Shvets}},\ }\href {\doibase
  10.1063/1.4775774} {\bibfield  {journal} {\bibinfo  {journal} {Phys.
  Plasmas}\ }\textbf {\bibinfo {volume} {20}},\ \bibinfo {pages} {013108}
  (\bibinfo {year} {2013})}\BibitemShut {NoStop}%
\bibitem [{\citenamefont {Vieira}, \citenamefont {Fonseca},\ and\ \citenamefont
  {Silva}(2016)}]{Vieira2016BubbleSummary}%
  \BibitemOpen
  \bibfield  {author} {\bibinfo {author} {\bibfnamefont {J.}~\bibnamefont
  {Vieira}}, \bibinfo {author} {\bibfnamefont {R.~A.}\ \bibnamefont {Fonseca}},
  \ and\ \bibinfo {author} {\bibfnamefont {L.~O.}\ \bibnamefont {Silva}},\
  }\href {\doibase 10.5170/CERN-2016-001.79} {\bibfield  {journal} {\bibinfo
  {journal} {CERN Yellow Rep.}\ }\textbf {\bibinfo {volume} {1}},\ \bibinfo
  {pages} {79} (\bibinfo {year} {2016})}\BibitemShut {NoStop}%
\bibitem [{\citenamefont {Golovanov}\ \emph
  {et~al.}(2016{\natexlab{b}})\citenamefont {Golovanov}, \citenamefont
  {Kostyukov}, \citenamefont {Thomas},\ and\ \citenamefont
  {Pukhov}}]{Golovanov_2016_PoP_23_093114}%
  \BibitemOpen
  \bibfield  {author} {\bibinfo {author} {\bibfnamefont {A.~A.}\ \bibnamefont
  {Golovanov}}, \bibinfo {author} {\bibfnamefont {I.~{\relax Yu}.}\
  \bibnamefont {Kostyukov}}, \bibinfo {author} {\bibfnamefont {J.}~\bibnamefont
  {Thomas}}, \ and\ \bibinfo {author} {\bibfnamefont {A.}~\bibnamefont
  {Pukhov}},\ }\href {\doibase 10.1063/1.4962565} {\bibfield  {journal}
  {\bibinfo  {journal} {Phys. Plasmas}\ }\textbf {\bibinfo {volume} {23}},\
  \bibinfo {pages} {093114} (\bibinfo {year} {2016}{\natexlab{b}})}\BibitemShut
  {NoStop}%
\bibitem [{Smi()}]{Smilei}%
  \BibitemOpen
  \href@noop {} {}\bibinfo {howpublished}
  {\url{https://smileipic.github.io/Smilei/}}\BibitemShut {NoStop}%
\bibitem [{\citenamefont {Derouillat}\ \emph {et~al.}(2017)\citenamefont
  {Derouillat}, \citenamefont {Beck}, \citenamefont {P{\'e}rez}, \citenamefont
  {Vinci}, \citenamefont {Chiaramello}, \citenamefont {Grassi}, \citenamefont
  {Fl{\'e}}, \citenamefont {Bouchard}, \citenamefont {Plotnikov}, \citenamefont
  {Aunai}, \citenamefont {Dargent}, \citenamefont {Riconda},\ and\
  \citenamefont {Grech}}]{Derouillat2017smilei}%
  \BibitemOpen
  \bibfield  {author} {\bibinfo {author} {\bibfnamefont {J.}~\bibnamefont
  {Derouillat}}, \bibinfo {author} {\bibfnamefont {A.}~\bibnamefont {Beck}},
  \bibinfo {author} {\bibfnamefont {F.}~\bibnamefont {P{\'e}rez}}, \bibinfo
  {author} {\bibfnamefont {T.}~\bibnamefont {Vinci}}, \bibinfo {author}
  {\bibfnamefont {M.}~\bibnamefont {Chiaramello}}, \bibinfo {author}
  {\bibfnamefont {A.}~\bibnamefont {Grassi}}, \bibinfo {author} {\bibfnamefont
  {M.}~\bibnamefont {Fl{\'e}}}, \bibinfo {author} {\bibfnamefont
  {G.}~\bibnamefont {Bouchard}}, \bibinfo {author} {\bibfnamefont
  {I.}~\bibnamefont {Plotnikov}}, \bibinfo {author} {\bibfnamefont
  {N.}~\bibnamefont {Aunai}}, \bibinfo {author} {\bibfnamefont
  {J.}~\bibnamefont {Dargent}}, \bibinfo {author} {\bibfnamefont
  {C.}~\bibnamefont {Riconda}}, \ and\ \bibinfo {author} {\bibfnamefont
  {M.}~\bibnamefont {Grech}},\ }\href@noop {} {\  (\bibinfo {year} {2017})},\
  \Eprint {http://arxiv.org/abs/1702.05128} {arXiv:1702.05128
  [physics.plasm-ph]} \BibitemShut {NoStop}%
\bibitem [{\citenamefont {Nerush}\ and\ \citenamefont
  {Kostyukov}(2010)}]{Nerush2010Quill}%
  \BibitemOpen
  \bibfield  {author} {\bibinfo {author} {\bibfnamefont {E.~N.}\ \bibnamefont
  {Nerush}}\ and\ \bibinfo {author} {\bibfnamefont {I.~{\relax Yu}.}\
  \bibnamefont {Kostyukov}},\ }\href@noop {} {\bibfield  {journal} {\bibinfo
  {journal} {Probl. Atom. Sci. Tech.}\ }\textbf {\bibinfo {volume} {4}},\
  \bibinfo {pages} {3} (\bibinfo {year} {2010})}\BibitemShut {NoStop}%
\bibitem [{\citenamefont {Godfrey}(1974)}]{godfrey1974numerical}%
  \BibitemOpen
  \bibfield  {author} {\bibinfo {author} {\bibfnamefont {B.~B.}\ \bibnamefont
  {Godfrey}},\ }\href {\doibase 10.1016/0021-9991(74)90076-X} {\bibfield
  {journal} {\bibinfo  {journal} {J. Comp. Phys.}\ }\textbf {\bibinfo {volume}
  {15}},\ \bibinfo {pages} {504} (\bibinfo {year} {1974})}\BibitemShut
  {NoStop}%
\bibitem [{\citenamefont {Esirkepov}, \citenamefont {Kato},\ and\ \citenamefont
  {Bulanov}(2008)}]{Esirkepov2008BowWave}%
  \BibitemOpen
  \bibfield  {author} {\bibinfo {author} {\bibfnamefont {T.~{\relax Zh}.}\
  \bibnamefont {Esirkepov}}, \bibinfo {author} {\bibfnamefont {Y.}~\bibnamefont
  {Kato}}, \ and\ \bibinfo {author} {\bibfnamefont {S.~V.}\ \bibnamefont
  {Bulanov}},\ }\href {\doibase 10.1103/PhysRevLett.101.265001} {\bibfield
  {journal} {\bibinfo  {journal} {Phys. Rev. Lett.}\ }\textbf {\bibinfo
  {volume} {101}},\ \bibinfo {pages} {265001} (\bibinfo {year}
  {2008})}\BibitemShut {NoStop}%
\bibitem [{\citenamefont {Luo}\ \emph {et~al.}(2016)\citenamefont {Luo},
  \citenamefont {Chen}, \citenamefont {Zhang}, \citenamefont {Yuan},
  \citenamefont {Yu}, \citenamefont {Shen}, \citenamefont {Yu}, \citenamefont
  {Weng}, \citenamefont {Schroeder},\ and\ \citenamefont
  {Esarey}}]{Luo2016waves}%
  \BibitemOpen
  \bibfield  {author} {\bibinfo {author} {\bibfnamefont {J.}~\bibnamefont
  {Luo}}, \bibinfo {author} {\bibfnamefont {M.}~\bibnamefont {Chen}}, \bibinfo
  {author} {\bibfnamefont {G.-B.}\ \bibnamefont {Zhang}}, \bibinfo {author}
  {\bibfnamefont {T.}~\bibnamefont {Yuan}}, \bibinfo {author} {\bibfnamefont
  {J.-Y.}\ \bibnamefont {Yu}}, \bibinfo {author} {\bibfnamefont {Z.-C.}\
  \bibnamefont {Shen}}, \bibinfo {author} {\bibfnamefont {L.-L.}\ \bibnamefont
  {Yu}}, \bibinfo {author} {\bibfnamefont {S.-M.}\ \bibnamefont {Weng}},
  \bibinfo {author} {\bibfnamefont {C.~B.}\ \bibnamefont {Schroeder}}, \ and\
  \bibinfo {author} {\bibfnamefont {E.}~\bibnamefont {Esarey}},\ }\href
  {\doibase 10.1063/1.4966047} {\bibfield  {journal} {\bibinfo  {journal}
  {Phys. Plasmas}\ }\textbf {\bibinfo {volume} {23}},\ \bibinfo {pages}
  {103112} (\bibinfo {year} {2016})}\BibitemShut {NoStop}%
\end{thebibliography}%

\end{document}